\documentclass[useAMS,usenatbib]{mn2e}
\usepackage{epsf,graphicx,rotating,url}

\title[Optical spectra of $\gamma$-ray bright blazars]{Optical spectroscopy of the $\gamma$-ray bright blazars PKS~0447$-$439 and PMN~J0630$-$24}

\author[H. Landt]{Hermine Landt$^1$\thanks{E-mail: hermine.landt@durham.ac.uk} \\ 
$^1$Department of Physics, University of Durham, South Road, Durham, DH1 3LE}

\begin{document}

\def\la{\mathrel{\hbox{\rlap{\hbox{\lower4pt\hbox{$\sim$}}}\hbox{$<$}}}}
\def\ga{\mathrel{\hbox{\rlap{\hbox{\lower4pt\hbox{$\sim$}}}\hbox{$>$}}}}

\font\sevenrm=cmr7
\def\MgII{Mg~{\sevenrm II}}
\def\FeII{Fe~{\sevenrm II}}

\date{Accepted ~~. Received ~~; in original form ~~}

\pagerange{\pageref{firstpage}--\pageref{lastpage}} \pubyear{2012}

\maketitle

\label{firstpage}

\begin{abstract}

The large majority of sources detected by the {\sl Fermi Gamma-ray
  Space Telescope} are blazars, belonging in particular to the blazar
subclass of BL Lacertae objects (BL Lacs). BL Lacs often have
featureless optical spectra, which make it difficult and sometimes
impossible to determine their redshifts. This presents a severe
impediment for the use of BL Lacs to measure the spectrum of the
extragalactic background light through its interaction with
high-energy $\gamma$-ray photons. I present here high-quality optical
spectroscopy of two of the brightest $\gamma$-ray blazars, namely,
PKS~0447$-$439 and PMN~J0630$-$24. The medium-resolution and high
signal-to-noise ratio optical spectra show clear absorption lines,
which place these BL Lacs at relatively high redshifts of $z\ge1.246$
for PKS~0447$-$439 and $z\ge1.238$ for PMN~J0630$-$24.
 
\end{abstract}


\section{Introduction}

Since its launch in 2008, the {\sl Fermi Gamma-ray Space Telescope}
has surveyed the sky with its Large Area Telescope
\citep[LAT;][]{FermiLAT}. The deep and uniform exposure, good
per-photon angular resolution and stable response of the LAT has
delivered the most sensitive and best-resolved all-sky survey to date
in the 100~MeV to 100~GeV energy range. A significant result delivered
by the {\sl Fermi Gamma-ray Space Telescope} was to show that the
extragalactic high-energy $\gamma$-ray sky is dominated by blazars. In
particular the blazar subclass of BL Lacertae objects (BL Lacs)
dominates the number counts in the recently released {\sl Fermi}
Second Source Catalog \citep{Fermi2}.

A major science goal for the {\sl Fermi} mission is to constrain the
near-infrared (near-IR) to ultraviolet (UV) spectrum of the
extragalactic background light (EBL) through its interaction with
high-energy $\gamma$-rays from extragalactic sources
\citep[e.g.,][]{FermiEBL}. The opacity of the Universe to high-energy
$\gamma$-rays can in principle be measured from the observed
$\gamma$-ray spectra of extragalactic sources if their intrinsic
$\gamma$-ray spectra are known. Therefore, this method requires
knowledge of the redshift of the extragalactic source. However, BL
Lacs usually have optical spectra with only weak both emission and
absorption features, which makes their redshift determination
difficult and often impossible \citep[e.g.,][]{L02, Sba05}.

Here I present results from high-quality optical spectroscopy that
allow a reliable redshift determination for two of the $\gamma$-ray
brightest BL Lacs in the southern sky, namely, PKS~0447$-$439 and
PMN~J0630$-$24 \citep{FermiBright}. Section 2 gives the details of the
optical spectroscopic observations, whereas I discuss the redshift
determination in Section 3. Section 4 presents the conclusions.

\section{The optical spectroscopy}

\begin{table*}
\caption{\label{log} 
Log of optical spectroscopy}
\begin{center}
\begin{tabular}{lllccrr}
\hline
Object Name & Observation date & Observatory & \multicolumn{2}{c}{Grism properties} & Exposure \\
&&& Dispersion & Range & (s) \\
&&& (\AA~pixel$^{-1}$) & (\AA) & \\
(1) & (2) & (3) & (4) & (5) & (6) \\
\hline
PKS~0447$-$439 & 2007 Jan 21 & NTT 3.6 m & 1.64 & 3680 -- 7036 &  900 \\
               & 2007 Mar 20 & CTIO 4 m  & 1.91 & 3360 -- 9450 &  900 \\
PMN~J0630$-$24 & 2007 Jan 21 & NTT 3.6 m & 1.64 & 3680 -- 7036 &  900 \\
               & 2007 Mar 21 & CTIO 4 m  & 1.91 & 3360 -- 9450 & 1800 \\
\hline
\end{tabular}
\end{center}
\end{table*}

\begin{figure*}
\centerline{
\includegraphics[angle=-90, scale=0.5]{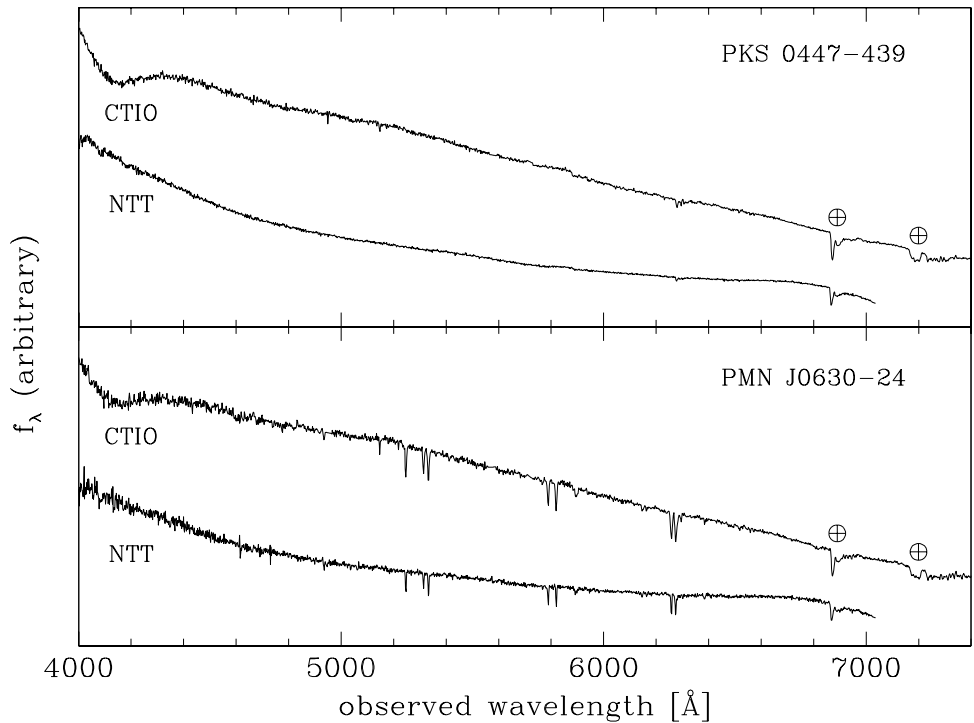}
}
\caption{\label{spectra} Optical spectra of the BL Lacs PKS~0447$-$439
  (upper panel) and PMN~J0630$-$24 (lower panel) shown in the
  observer's frame. In each panel the upper and lower spectrum are
  from the CTIO 4 m and NTT 3.6 m telescopes, respectively. Telluric
  absorption bands are indicated by circled plus signs.}
\end{figure*}

The sources PKS~0447$-$439 and PMN~J0630$-$24 are part of the Deep
X-ray Blazar Survey \citep[DXRBS;][]{Per98, L01} within which they
were classified as BL Lacs without a reliable redshift. For both
sources \citet{L08c} presented deep, high angular resolution radio
maps that provided accurate source positions. Based on these radio
observations it is clear that the optical finder and optical spectrum
of the source PKS~0447$-$439 presented by \citet{Craig97}, whose
redshift of $z=0.107$ is currently listed in the NASA/IPAC
Extragalactic Database (NED), must be misidentifications.

In order to improve on the redshift determination of all featureless
DXRBS BL Lacs proposals for high-quality optical spectroscopy were
successfully submitted to the Magellan 6.5 m, Cerro Tololo
Inter-American Observatory (CTIO) 4 m and European Southern
Observatory (ESO) New Technology Telescope (NTT) 3.6 m telescopes. The
BL Lacs PKS~0447$-$439 and PMN~J0630$-$24 were observed at both the
CTIO 4 m and NTT 3.6 m observatories and Table \ref{log} gives the
details of the observational set-up. Since any emission or absorption
features in the spectra of these BL Lacs were expected to be weak, the
observational strategy aimed at obtaining data of both a medium
spectral resolution and a relatively high signal-to-noise ratio
(S/N~$\ga 50$).

The long-slit spectra were reduced using standard routines from the
IRAF software package. In particular the 2-dimensional spectral files
were trimmed, overscan- and bias-subtracted, normalized, rectified and
wavelength-calibrated. Subsequently the spectrum was extracted and the
1-dimensional spectral file flux-calibrated using photometric standard
stars observed the same night. The final spectra were corrected for
Galactic extinction using the IRAF task \mbox{\sl onedspec.deredden}
with input $A_{\rm V}$ values derived from Galactic hydrogen column
densities published by \citet{DL90}. The results are shown in
Fig. \ref{spectra}.

The final average signal-to-noise ratio of all spectra is S/N~$\sim
80$. The CTIO spectra were obtained under good seeing conditions,
whereas patchy cloud was present during the NTT night. The observed
flux at $\lambda = 5500$~\AA~in the CTIO spectra is
4.39$\times$10$^{-15}$ erg~s$^{-1}$~cm$^{-2}$~\AA$^{-1}$ for
PKS~0447$-$439 and 1.18$\times$10$^{-15}$
erg~s$^{-1}$~cm$^{-2}$~\AA$^{-1}$ for PMN~J0630$-$24.  I note that the
bump and spectral upturn visible in the CTIO spectra at an observed
wavelength of $\lambda \la 4200$~\AA~and the slight spectral downturn
visible in the NTT spectra at $\lambda \ga 7000$~\AA~are due to the
poor spectral response of the grisms in these regions.

\section{Redshift determination}

\begin{figure}
\centerline{
\includegraphics[scale=0.42]{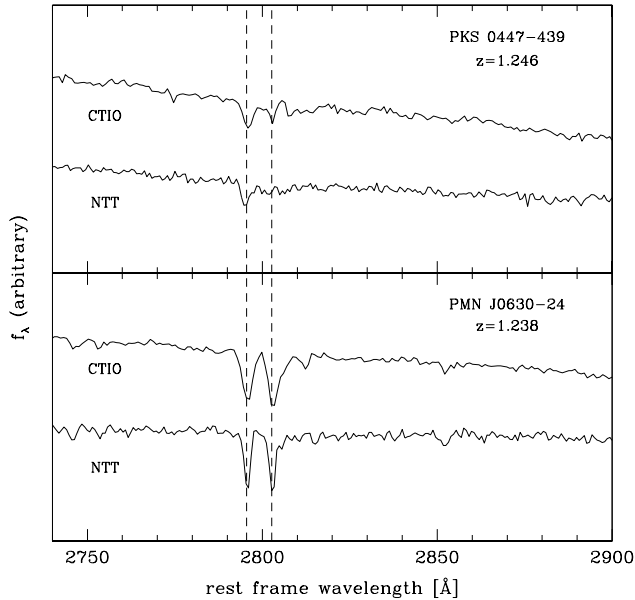}
}
\caption{\label{zoomspectra} Optical spectra of the BL Lacs
  PKS~0447$-$439 (upper panel) and PMN~J0630$-$24 (lower panel) shown
  in the rest frame. Based on the \MgII~$\lambda 2800$ doublet in
  absorption (marked by the vertical dashed lines) detected in both
  the CTIO 4~m and NTT 3.6~m spectra the redshift for PKS~0447$-$439
  is $z\ge1.246$ and that of PMN~J0630$-$24 is $z\ge1.238$.}
\end{figure}

The optical spectra of the two BL Lacs PKS~0447$-$439 and
PMN~J0630$-$24 show clear absorption features but no emission
lines. The absorption features are prominent and numerous in the
spectra of the source PMN~J0630$-$24, whereas only weak absorption
features are detected in the spectra of the source
PKS~0447$-$439. Based on the \MgII~$\lambda 2800$ doublet in
absorption, which is clearly detected in all spectra (see
Fig. \ref{zoomspectra}), the redshift for the source PKS~0447$-$439 is
$z\ge1.246$ and that of the source PMN~J0630$-$24 is $z\ge1.238$. The
observed \MgII~equivalent widths are $W_{\lambda}=0.3$~\AA~and
1.4~\AA~for PKS~0447$-$439 and PMN~J0630$-$24, respectively. Other
absorption lines detected in the spectra of PMN~J0630$-$24 are
\FeII~lines, most notably \FeII~$\lambda 2382$, \FeII~$\lambda 2586$
and \FeII~$\lambda 2600$.

Due to the lack of spectroscopic redshifts in numerous BL Lacs several
authors have investigated alternative methods to estimate this
quantity. The method of \citet{Pir07} uses the fact that BL Lacs are
hosted by luminous ellipticals of almost constant luminosity
\citep[e.g.,][]{Wur96, Urry00}. Then, assuming a plausible lower limit
for the jet/galaxy ratio of a featureless BL Lac one can derive a
lower limit on the redshift from the observed optical $V$
magnitude. Based on this method \citet{L08c} estimated a lower limit
on the redshift of the source PKS~0447$-$439 of $z>0.176$, assuming a
jet/galaxy ratio of ten. Clearly, this lower limit strongly
underestimates the true redshift of this BL Lac.

Similarly, \citet{Pra11} have presented an estimate of the redshift of
the source PKS~0447$-$439 of $z=0.2\pm0.05$. Their method assumes that
the spectral slopes of the {\sl Fermi} LAT $\gamma$-ray spectrum and
that measured at very high $\gamma$-ray energies by ground-based
Cherenkov telescopes are the same. Then, any observed differences
between the two $\gamma$-ray spectra are attributed to the interaction
with the EBL photons, which will be stronger at higher energies and
will depend on the redshift of the source. However, the major unknown
quantity in the application of this method is the spectrum of the EBL.

Recently, \citet{Rau12} have introduced a method to estimate the
redshift of a featureless BL Lac based on the fact that absorption by
neutral hydrogen along the line of sight will cause a prominent
absorption feature in the broad-band spectrum. Then, a power-law fit
to the near-IR to UV (quasi-simultaneous) spectral energy distribution
of a BL Lac can place an upper limit for low-redshift sources and give
an estimate for high-redshift sources. These authors obtained for the
source PMN~J0630$-$24 a redshift of $z=1.60^{+0.10}_{-0.05}$, which is
above the value obtained from optical spectroscopy. However, since no
emission lines are detected in the optical spectrum of the source
PMN~J0630$-$24, it cannot be excluded that the observed absorption
lines are due to an intervening system rather than the host galaxy.

\section{Conclusions}

I have presented high-quality (medium-resolution, high signal-to-noise
ratio) optical spectroscopy of two of the $\gamma$-ray brightest BL
Lacs, namely, PKS~0447$-$439 and PMN~J0630$-$24. Based on a clear
detection of the \MgII~$\lambda 2800$ doublet in absorption the
redshift for the source PKS~0447$-$439 is $z\ge1.246$ and that of the
source PMN~J0630$-$24 is $z\ge1.238$. Alternative methods proposed so
far to estimate the redshifts of featureless BL Lacs give in the case
of PKS~0447$-$439 a relatively low value of $z\sim0.2$ and in the case
of PMN~J0630$-$24 an overestimate of $z\sim1.6$. Therefore,
high-quality optical spectroscopy that allows the detection of even
the weakest features in the spectra of BL Lacs remains an absolute
necessity.

\section*{Acknowledgments}

I thank Eric S. Perlman and Paolo Padovani for their help with the
observations. I acknowledge financial support by the European Union
through the COFUND scheme. This research has made use of the NASA/IPAC
Extragalactic Database (NED), which is operated by the Jet Propulsion
Laboratory, California Institute of Technology, under contract with
the National Aeronautics Space Administration.

\bibliography{/Users/herminelandt/references}

\bsp
\label{lastpage}

\end{document}